\documentclass[fleqn,10pt]{wlscirep}
\usepackage[utf8]{inputenc}
\usepackage[T1]{fontenc}

\usepackage{graphicx}% Include figure files
\usepackage{dcolumn}% Align table columns on decimal point
\usepackage{bm}% bold math
\usepackage[subrefformat=parens]{subcaption}
\usepackage{caption}

\title{Daily-activity-dependency of effective reproduction number in COVID-19 pandemic: direct modelling from GPS data}

\author[1]{Jun'ichi Ozaki}
\author[2]{Yohei Shida}
\author[1,3]{Hideki Takayasu}
\author[1,2,*]{Misako Takayasu}
\affil[1]{Institute of Innovative Research, Tokyo Institute of Technology, 4259 Nagatsuta-cho, Midori-ku, Yokohama 226-8503, Japan}
\affil[2]{Department of Mathematical and Computing Science, School of Computing, Tokyo Institute of Technology, 4259 Nagatsuta-cho, Midori-ku, Yokohama 226-8503, Japan}
\affil[3]{Sony Computer Science Laboratories, Inc., 3-14-13, Higashigotanda, Shinagawa-ku, Tokyo 141-0022, Japan}

\affil[*]{Electronic Address: takayasu.m.aa@m.titech.ac.jp}

\begin{abstract}
During the COVID-19 pandemic, governments faced difficulties in implementing mobility restriction measures, as no clear quantitative relationship between human mobility and infection spread in large cities is known. We developed a model that enables quantitative estimations of the infection risk for individual places and activities by using smartphone GPS data for the Tokyo metropolitan area. The effective reproduction number is directly calculated from the number of infectious social contacts defined by the square of the population density at each location. The difference in the infection rate of daily activities is considered, where the `stay-out' activity, staying at someplace neither home nor workplace, is more than 28 times larger than other activities. Also, the contribution to the infection strongly depends on location. We imply that the effective reproduction number is sufficiently suppressed if the highest-risk locations or activities are restricted. We also discuss the effects of the Delta variant and vaccination.

\end{abstract}

\begin{document}

\flushbottom
\maketitle
\thispagestyle{empty}

\section*{Introduction}

Since the beginning of the COVID-19 pandemic in 2019, there have been 452 million confirmed cases and over six million deaths globally as of 12 March 2022, posing serious healthcare challenges \cite{Covid:General01,Covid:General02,Covid:General03}. Most governments have struggled to control this disease and simultaneously minimize its damage to daily and economic activities owing to limited time and resources \cite{Covid:CounterMeasures01,Gov:Tokyo01,Covid:Damage01,Covid:Damage02,Covid:Damage03}. Along with vaccinations, nonpharmaceutical interventions are considered essential for managing this disease \cite{Covid:NPITokyo01,Covid:NPI01,Covid:NPI02,Covid:NPI03,Covid:NPI04}. For example, the governments around the Tokyo metropolitan area in Japan declared states of emergency (SoEs) that limited daily and economic activities including schools, department stores, cinemas, restaurants, bars, and travel to reduce human mobility in public spaces \cite{Gov:Tokyo01,Covid:NPITokyo01}. Consequently, the number of social contacts and, thereby, the effective reproduction number decreased. Governments require reliable quantitative estimations of the effect of such policies on the pandemic, that is, a predictable model, for making informed decisions.

Studies have estimated the effectiveness of lockdowns or non-compulsory measures such as SoEs on reducing human mobility \cite{Covid:NPITokyo01,Covid:NPI01,Covid:NPI02,Covid:NPI03,Covid:NPI04}. However, the effect of reduced human mobility on the pandemic remains unclear, as it is insufficient to investigate the number of social contacts alone. Infectious diseases, including the COVID-19 pandemic, have been studied using global positioning system (GPS) data \cite{Covid:GPS01,Covid:GPS02,Covid:GPS03,Covid:GPS04,Covid:GPS05,Covid:GPS06,Covid:GPS07,Covid:GPS08,Covid:GPS09,Covid:GPS10}. However, with such diseases, the types of social contacts are considered more critical. Specifically, the infection rate is known to strongly depend on whether social contacts are wearing masks or talking and on the ventilation state of the rooms they are in \cite{Covid:ConditionDependency01,Covid:ConditionDependency02,Covid:ConditionDependency03}. The point is that the number of social contacts through each daily activity type should be investigated.

In this study, we propose a model to predict the effective reproduction number of COVID-19 based on daily human activities inferred from smartphone GPS data. From these data, first, we categorize citizens' daily activity patterns into four types and estimate the population density for each activity, location, and time in the Tokyo metropolitan area. We also calculate the number of social contacts for each activity by assuming it to be proportional to the square sum of the population density. Second, we propose an activity-dependent infection model based on the susceptible-infectious-removed (SIR) model. This model seems to have a lower resolution than that of other compartmental models (e.g., susceptible-exposed-infectious-removed (SEIR) model). However, it is suitable for direct formulation using GPS data. The effective reproduction number is a linear combination of the number of social contacts, and the coefficients are the infection rate per contact. We determine the parameters to fit the empirical data before the spread of the Delta variant and verify that the model is sufficient to predict the effective reproduction number. Then, we compare the parameters to observe the activity that has the highest infection risk. We also calculate the effective reproduction number for each daily activity, location, and individual. The model prediction is valid for up to around two weeks after the human mobility data are obtained. Third, we investigate effects other than those of human mobility. We show that the domination of the Delta variant is described by only two parameters: starting date and ratio of effective reproduction number to that of existing variants. Furthermore, we demonstrate the effectiveness of vaccination through its high prevention ratio of infection in a metropolis.

\section*{Results}

\subsection*{Epidemiological data}

\begin{figure}[htbp]
%    \begin{minipage}[b]{0.5\linewidth}
        \centering
            \includegraphics[width=8cm]{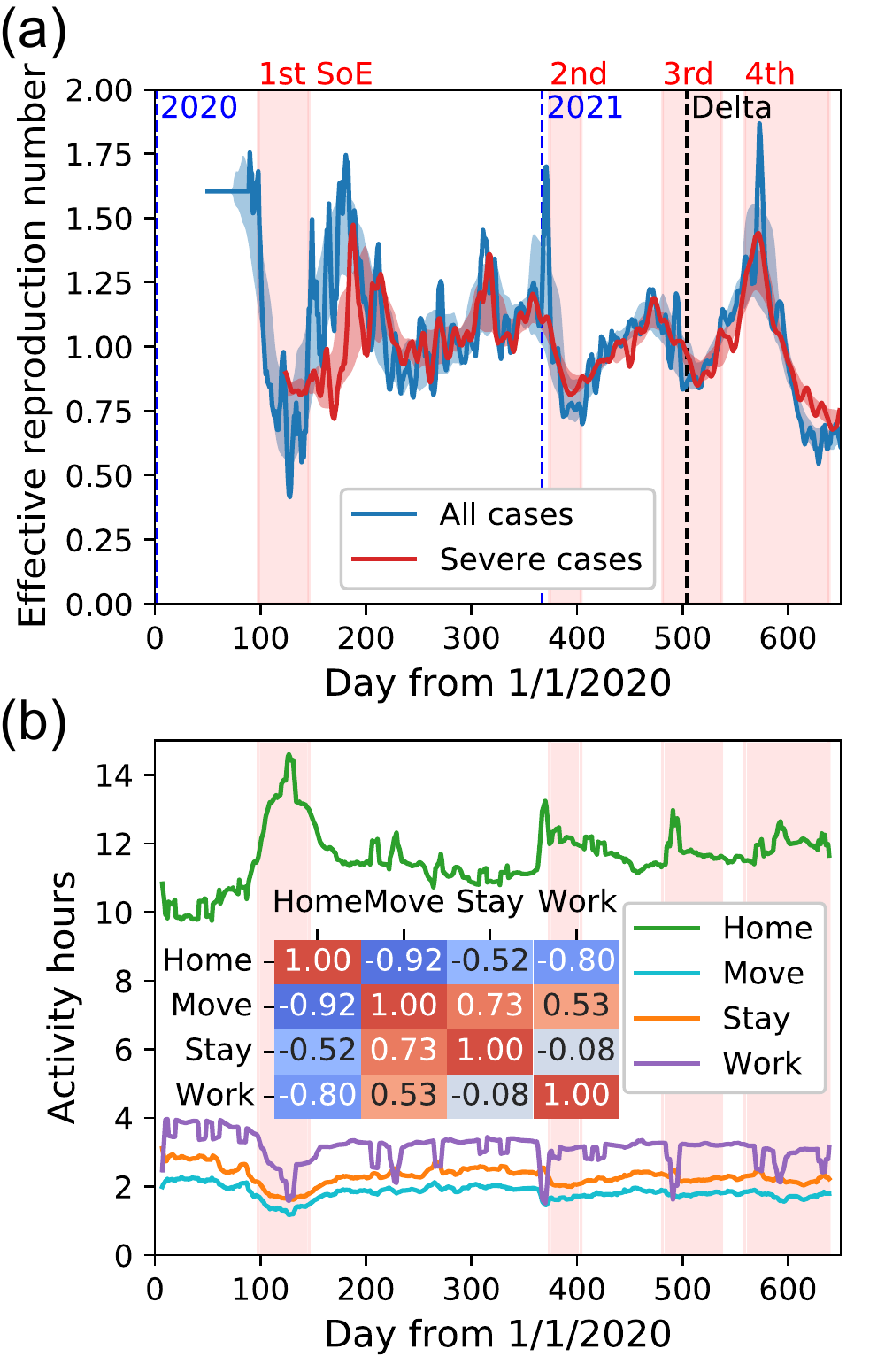}
        \caption{ (a) Effective reproduction number of all infections and severe cases in Tokyo metropolitan area. The SoEs and beginnings of years are also shown. The first case of the Delta variant in Japan is represented as `Delta'. Severe cases are plotted taking into account the 12-day delay. The light colored range is estimated as follows. The centre of the range is the 31-day moving average of the original time series, and the range width is the 31-day moving average of the absolute difference between the centre and original time series. Before 30 Mar. 2020, the effective reproduction number is estimated by the average of 40 days: the ratio of the number in the first 20 days to that in the last 20 days to the power of 1/4 because of the lack of data reliability: the number of PCR tests per day is under 10,000 in this period.
(b) Time series of mean duration per day of each activity in the Tokyo metropolitan area. The 7-day moving average is taken. The 1st--4th SoEs are also shown. The Pearson's correlation among the mean duration of the four activities in the span $200 \leq t < 500$ is calculated.}
\label{fig:effective_reproduction_number_TimeSeriesFourStates}
%\end{minipage}
\end{figure}

First, we analysed COVID-19 epidemiological data in the Tokyo metropolitan area from 1 Feb. 2020 to 31 Oct. 2021 \cite{Gov:JapanMHLW}. $t$ denotes the day count from the beginning of 2020 (e.g., $t=1$ means 1 Jan. 2020) and $I^{\mathrm{new}}(t)$, the number of new COVID-19 infection cases on day $t$. The effective reproduction number $R_e^{\mathrm{data}}(t)$ was estimated as \cite{Covid:CislaghiMethod}
\begin{equation}
    R_e^{\mathrm{data}}(t)=\frac{I^{\mathrm{new}}(t+\gamma^{-1})}{I^{\mathrm{new}}(t)},
\end{equation}
where $\gamma^{-1} = 5$ is the mean generation time \cite{Covid:SerialInterval01}. Fig. \ref{fig:effective_reproduction_number_TimeSeriesFourStates}(a) shows a plot of the effective reproduction number. Similarly, we plotted the effective reproduction number limited in severe cases, which is defined using only the number of new severe cases instead of the number of all infections. The correlation between the effective reproduction numbers for all cases and severe cases is maximized if that for severe cases is regarded to be delayed by 12 days. Both are consistent with each other after $t=200$. The 1st to 4th SoEs in the Tokyo metropolitan area \cite{Gov:Tokyo01} are also shown. During each SoE, the effective reproduction number started to decrease after around two weeks. The Delta variant was first detected in Japan on 18 May 2021 (i.e., $t=504$) \cite{Covid:FirstDelta01}.

\subsection*{Human mobility and activity patterns}

We analysed empirical human mobility data in Japan, especially in the Tokyo metropolitan area, from 1 Jan. 2020 to 30 Sep. 2021 \cite{Agoop}. We calculated the population density at each time and location to estimate the number of infection routes. The infection probability of each infection route depends on how people contact each other and, consequently, on the activity. Therefore, we categorized the GPS user activity into four states, staying home (`home'), moving (`move'), staying out (`stay'), and working (`work'), to distinguish the activity-dependent infection probability with a time resolution of 15 min. The definition is as follows: (1) home is the state when the user is within the 1 km-square corresponding to their home, (b) work is the state when the user is within the 1 km-square corresponding to their workplace, (c) move is the state in which the user is within a 1 km-square that is different from the one in which they were 15 min ago and that does not correspond to either their home or workplace, and (d) stay-out is any other situation, e.g., in restaurants, department stores, or stadiums away from their home. 
Here, the home and work regions are estimated from the home/work city data in the GPS dataset, as explained in the \textit{Methods} section.
Fig. \ref{fig:effective_reproduction_number_TimeSeriesFourStates}(b) shows a plot of the time series of the mean duration per day of the four activities, where we take the 7-day moving average for avoiding the periodicity related to weekdays. The total time per day is 19 h because the data point is from 05:00 to 24:00. During the pandemic, roughly speaking on average, users were at home for 12 h (not including midnight), at work for 3 h, moved for 2 h, and stayed out for 2 h. Note that this is averaged over all days including holidays for all users. We also calculated the correlation among the mean durations of the four states in the span $200 \leq t < 500$. This span was used later for model parameter fitting. Move was strongly correlated with work and stay.

\subsection*{Model}
We derived an infection model based on the classical SIR model using the GPS data for the four states with several assumptions. The classical SIR model is given by the following set of equations:
\begin{eqnarray}
    \frac{dS(t)}{dt}&=&-\frac{\beta}{N}S(t)I(t)\\
    \frac{dI(t)}{dt}&=&\frac{\beta}{N}S(t)I(t)-\gamma I(t)\\
    \frac{dR(t)}{dt}&=&\gamma I(t),
\end{eqnarray}
where $t$ is the time; $S(t)$, the number of susceptible people; $I(t)$, the number of infectious people; $R(t)$, the number of removed people (either recovered, quarantined, or dead); and $N=S(t)+I(t)+R(t)$, the total number. 
The parameter $\beta$ represents the strength of infection spread, and the parameter $\gamma$ determines the timescale from infected state to recovered or quarantined state. The reciprocal of $\gamma$ is the mean generation time \cite{Covid:SerialInterval01}: $\gamma^{-1}=5$. 
We adopted the following four assumptions: \\
(A1) the number of removed people is much smaller than the number of susceptible people, \\
(A2) no nonlocal infections occur between different spaces or time periods, \\
(A3) infectious people are distributed uniformly in the target area (i.e., Tokyo metropolitan area), and \\
(A4) no infections occur between different activities. \\
Assumption (A1) comes from the low cumulative number of COVID-19 cases in Japan (below 2\% on 30 Oct. 2021) \cite{Gov:JapanMHLW}. Then, the number of susceptible people is assumed to be constant, that is, the whole population of Japan. Assumption (A2) means that we discard indirect infections such as droplet infection with long-distance or contact infection after a long time \cite{Covid:InfectionRange01}. Nonetheless, indirect infections are effectively included if the population in the target area does not change drastically. Assumption (A3) is used for simplifying the model. 
As our GPS data does not contain users' privacy including the information of infection, we simply assume that infected people distribute uniformly. 
%Infected people finally go home, and the distribution is assumed to be uniform because of the lack of information. 
%Human mobility must be limited in a given area such as the Tokyo metropolitan area. 
We discard the effect from infected people who went outside or inside of the Tokyo metropolitan area. 
%Assumption (A4) comes from the extreme ratio between workers and customers in typical daily situations: fewer social contacts occur between workers and customers than among workers and among customers. It limits the number of model parameters; without this assumption, 16 parameters will be considered.
Assumption (A4) is the intuition that social contacts between different activities are fewer than those within the same activity. For example, in a restaurant, `work' and `stay' people will have much fewer contacts than between `stay' and `stay'.

By applying the first assumption (A1), we only need to consider the second equation of the SIR model because infection causes a negligible change in $S(t)$. In this case, the equation becomes
\begin{equation}
    \frac{dI(t)}{dt}=\gamma\left(R_e(t)-1\right)I(t),
\end{equation}
where $R_e(t)=\frac{\beta}{N\gamma}S(t)$ is an effective reproduction number. 
We discretize the equation by setting $dt=\gamma^{-1}$ as
\begin{equation}
    I(t+\gamma^{-1})=\tilde{\beta}S(t) I(t), 
\end{equation}
where $\tilde{\beta} = \frac{\beta}{N\gamma}$. This discretization is needed for fitting the empirical data.

Next, we adopt the second assumption (A2) of local infections. We suppose that all people in the target area have equal possibility of close contact with each other. In this case, the time evolution of $I(t)$ is described by the product of $I(t)$ and $S(t)$ because the number of infection routes or social contacts is proportional to $I(t)S(t)$. In this sense, the classical SIR model is exact only if the population density is uniform. However, this is not the case in real situations. 
Therefore, we divide the area into squares (grids) in which social contacts are equally possible among all people in a square. The size of the squares is set to 1 km$\times$1 km. The case of 100 m$\times$100 m \cite{Covid:NPITokyo01} is discussed in Supplementary note 1. Thus, we obtain
\begin{equation}
    I_{m\tau}(t+\gamma^{-1})=\tilde{\beta}S_{m\tau}(t) I_{m\tau}(t), 
\end{equation}
where $m$ is the square label, and $\tau$ is the time label in the considered area $A$ and date $t$. The area $A$ is the Tokyo metropolitan area, and it contains 36,898 1 km-squares. The time label $\tau$ moves in 15 min intervals around the current date $t$ for five days, from $t-2$ to $t+2$, according to the discretization unit $\gamma^{-1}=5$. The parameter $\tilde{\beta}$ takes a different value from the earlier equations because the space size and period are different: $\tilde{\beta}$ is multiplied by (number of 1 km-squares)/(number of time steps) if the density is uniform in space and time.

The third assumption (A3) gives the following relation:
\begin{equation}
    I_{m\tau}(t)=\frac{S_{m\tau}(t)}{S(t)}I(t),
\end{equation}
where $S(t)=\sum_{m\in A}S_{m\tau}(t)$, $I(t)=\sum_{m\in A}I_{m\tau}(t)$, and $A$ is the target area. 
Therefore, the effective reproduction number is given as
\begin{equation}
    R_e(t)=
    \sum_{m\in A,\tau\in[t-2,t+2]}\tilde{\beta}\frac{\left(S_{m\tau}(t)\right)^2}{S(t)}. 
\end{equation}

Finally, assumption (A4) is applied. We divide the population density $S_{m\tau}(t)$ into that of the activities $S_{am\tau}(t)$ and introduce activity-dependent parameters $\beta_a$. The equation is 
\begin{equation}
    R_e(t)=\frac{1}{S(t)}\sum_{a}\beta_a \sum_{m\in A,\tau\in[t-2,t+2]}\left(S_{a m\tau}(t)\right)^2,
\end{equation}
where the time label $\tau$ moves from $t-2$ to $t+2$, and the activity label $a$ takes the values `home', `move', `stay', and `work'. The coefficient $\beta_a$ is supposed to be constant in the timespan considered.
For simplicity, we approximate the sum as follows:
\begin{equation}
    \sum_{\tau\in[t-2,t+2]}\left(S_{a m\tau}(t)\right)^2 \simeq 5 \sum_{\tau\in t}\left(S_{a m\tau}(t)\right)^2,
\end{equation}
for the equation to have values only on date $t$, where the time label $\tau$ moves only on date $t$ in the right-hand side. Thus, the effective reproduction number is a linear combination of the human mobility time series,
\begin{equation} \label{Rp_by_Moment}
    R_e(t)=\frac{1}{S(t)}\sum_{a}\beta_a M_a(t),
\end{equation}
by introducing $M_a(t)=\gamma^{-1}\sum_{m\in A,\tau\in t}\left(S_{a m\tau}(t)\right)^2$, the population moment of activity $a$ on date $t$. We emphasize that the effective reproduction number is simply determined only by the human mobility on the same day. The parameters $\beta_a$ are interpreted as the infection rate during activity $a$ per infection route and 15 min period.

%\subsection*{Parameter estimation}

\begin{figure}[htbp]
%    \begin{minipage}[b]{0.5\linewidth}
        \centering
            \includegraphics[width=8cm]{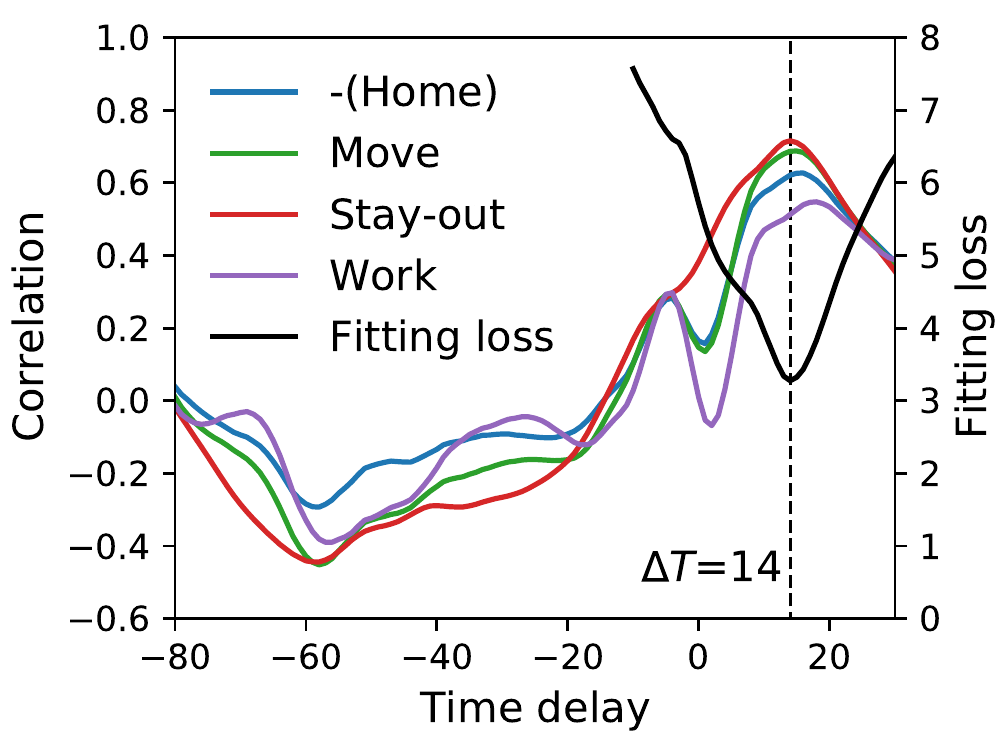}
        \caption{Correlation between the effective reproduction number and the population moments time series delayed by $\Delta T$, and fitting loss to the effective reproduction number by the population moments as a function of the delay $\Delta T$ in the period $200 \leq t < 500$. The fitting loss is calculated as the residual of the square sum in the log space. For the correlation, the peak and trough are observed at around $\Delta T = 14$ and $\Delta T = -60$, respectively. For the fitting loss, it is minimized at $\Delta T = 14$. }
        \label{fig:TimeDelayEstimation}
%    \end{minipage}
\end{figure}

We determine the model parameters $\beta_a$ to fit the effective reproduction number; however, the observed effective reproduction number is based on the report date and not the infection date. We have to consider a typical time delay from infection to report $\Delta T$ to deal with it. 
In light of this effect, Eq. (\ref{Rp_by_Moment}) is modified as follows:
\begin{equation}
    R_e^{\mathrm{data}}(t+\Delta T)=\frac{1}{S(t)}\sum_{a}\beta_a M_a(t). 
\end{equation}
Here, we use the 7-day moving average of $M_a(t)$ to remove the periodicity of weekdays as well as the definition of the empirical effective reproduction number. We estimate the time delay from infection to report $\Delta T$ in the period $200 \leq t < 500$. Figure \ref{fig:TimeDelayEstimation} shows the correlation between the effective reproduction number and the population moment of each activity delayed by $\Delta T$. The peak at $\Delta T \sim 14$ is the delay of the infection reports, whereas the trough at $\Delta T \sim -60$ means that human mobility is decreased after infection spread. The optimal value $\Delta T=14$ is derived by minimizing a fitting loss to the effective reproduction number by the population moments as a function of $\Delta T$, where the fitting loss is the squared distance in the log-10 space, to prevent the estimation from being dominated only by the significant value of the effective reproduction number. This value is consistent with that reported in a previous study \cite{Covid:Periods01,Covid:Periods02}. 

%The population moment defined by 100m-square population $S_{am\tau}(t)$ is approximated by 1km-population $S_{am_\mathrm{1km}\tau}(t)$ (see \textit{Methods}), and t

The remaining problem is the multicollinearity of the move population moment; the correlation of the population moments is 0.89 between move and stay-out and 0.90 between move and work. To address this, we assume that the coefficient $\beta_a$ for move and work is the same because the situations in these two activities are similar: people wear masks in trains (move) and offices (work) but not always in restaurants (stay-out). 

We fit the model parameters to the data for all cases in the period $200 \leq t < 500$. We do not use data for $t<200$ because the social situation drastically changed during the 1st SoE (e.g., mask distribution in Japan\cite{News:MaskDistribution01}), and the data are not stationary. Furthermore, the effective reproduction numbers for all cases and severe cases are different: epidemiological data are not reliable in that span. The fitting is performed under the log space to prevent the estimation from being dominated only by the significant value of $R_e(t)$. As a result, the parameters are $\beta_\mathrm{home}=(1.2 \pm 0.1)\times 10^{-7}$, $\beta_\mathrm{move}=\beta_\mathrm{work}=(6.2 \pm 2.8)\times 10^{-8}$, and $\beta_\mathrm{stay}=(3.4 \pm 0.2)\times 10^{-6}$, where the unit is per 15 min. Figure \ref{fig:effective_reproduction_number_fit_DV} shows that the model explains the data in the period; however, they are different before $t=200$ and after $t=500$. Before $t=200$, the effective reproduction number for severe cases is more consistent with the model result. After $t=500$, the effects of the Delta variant and the vaccination result in differences \cite{Covid:FirstDelta01,Gov:JapanKantei01}. Thus, the classical SIR model is suitable for direct GPS data modelling. Despite its simplicity, we observe that the following model quantitatively explains the COVID-19 epidemic.

\subsubsection*{Delta variant and vaccination effects}

\begin{figure}[htbp]
%    \begin{minipage}[b]{0.5\linewidth}
        \centering
            \includegraphics[width=8cm]{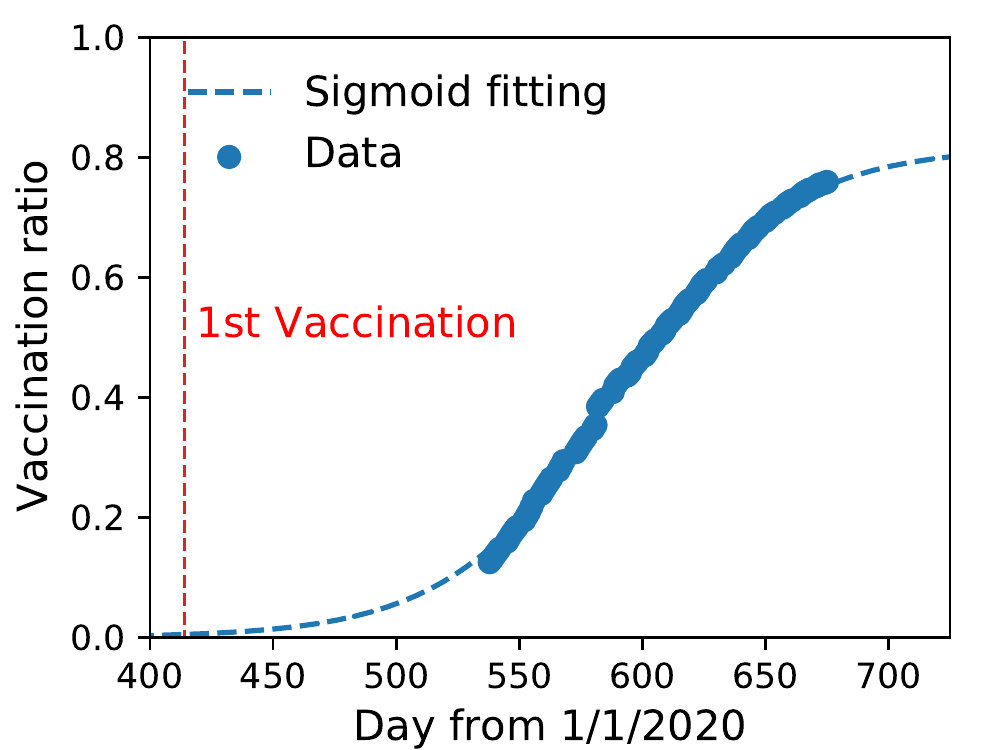}
        \caption{Vaccination ratio of the population of Japan calculated as $R_V(t)=$(number of vaccinations in Japan up to date $t$)/2(Japanese population). The first vaccination day is shown. A sigmoid-like function $R_V^\infty S(\gamma_V(t-t_V))$ fits the data, and the parameters are $t_V=590$, $\gamma_V^{-1}=34.6$, and $R_V^\infty=0.817$.}
        \label{fig:VaccinationRatio}
%    \end{minipage}
\end{figure}

The effects of the Delta variant and vaccination are estimated as follows \cite{Covid:DeltaRt02}. 
Let $r_\delta = R^\delta / R^0$ be the ratio of the effective reproduction number of the Delta variant $R^\delta$ to that of the other existing variants $R^0$. Furthermore, we assume that the numbers of people infected by the Delta variant $I^\delta(t)$ and the existing variants $I^0(t)$ are the same at $t=t_\delta$. If both variants increase independently, the ratio is $I^\delta(t)/I^0(t) = (R^\delta)^{\gamma(t-t_\delta)}/(R^0)^{\gamma(t-t_\delta)} = r_\delta^{\gamma(t-t_\delta)}$. 
By definition, the effective reproduction number of mixed variants $R^{\mathrm{mix}}(t)$ is
\begin{equation}
    R^{\mathrm{mix}}(t)/R^0=\frac{1}{R^0}\frac{R ^\delta I^\delta(t)+R^0 I^0(t)}{I^\delta(t)+I^0(t)}=1+\frac{R ^\delta-R_0}{R^0}\frac{ I^\delta(t)}{I^\delta(t)+I^0(t)}=1+(r_\delta-1)S(\gamma\log(r_\delta)\cdot(t-t_\delta)),
\end{equation}
where $S(x)=1/(1+\exp(-x))$ is the standard sigmoid function. The range of $R^{\mathrm{mix}}(t)$ is $R^0$ to $R^\delta$. 
Therefore, the effect of the Delta variant is introduced by multiplying by the factor $1+(r_\delta-1)S(\gamma\log(r_\delta)\cdot(t-t_\delta))$. 
The vaccination model is also multiplied by a factor
\begin{equation}
    1 - C_V R_V(t),
\end{equation}
where $C_V$ is the infection prevention ratio of the vaccine \cite{Covid:VacEff02,Covid:VacEff03}, and $R_V(t)$ is the vaccination ratio in the target area. We approximate the vaccination ratio as $R_V(t)=$(number of vaccinations in Japan up to date $t$)/2(Japanese population) and fit its data by a sigmoid function 
\begin{equation}
    R_V(t)=R_V^\infty S(\gamma_V(t-t_V)), 
\end{equation}
as shown in Fig. \ref{fig:VaccinationRatio}. The parameters are $t_V=590$, $\gamma_V^{-1}=34.6$, and $R_V^\infty=0.817$. The sigmoid function $S(x)$ is suitable for fitting the vaccination ratio because it is saturated in the limit $x\to\pm\infty$. 

\begin{figure}[htbp]
%    \begin{minipage}[b]{0.5\linewidth}
        \centering
            \includegraphics[width=8cm]{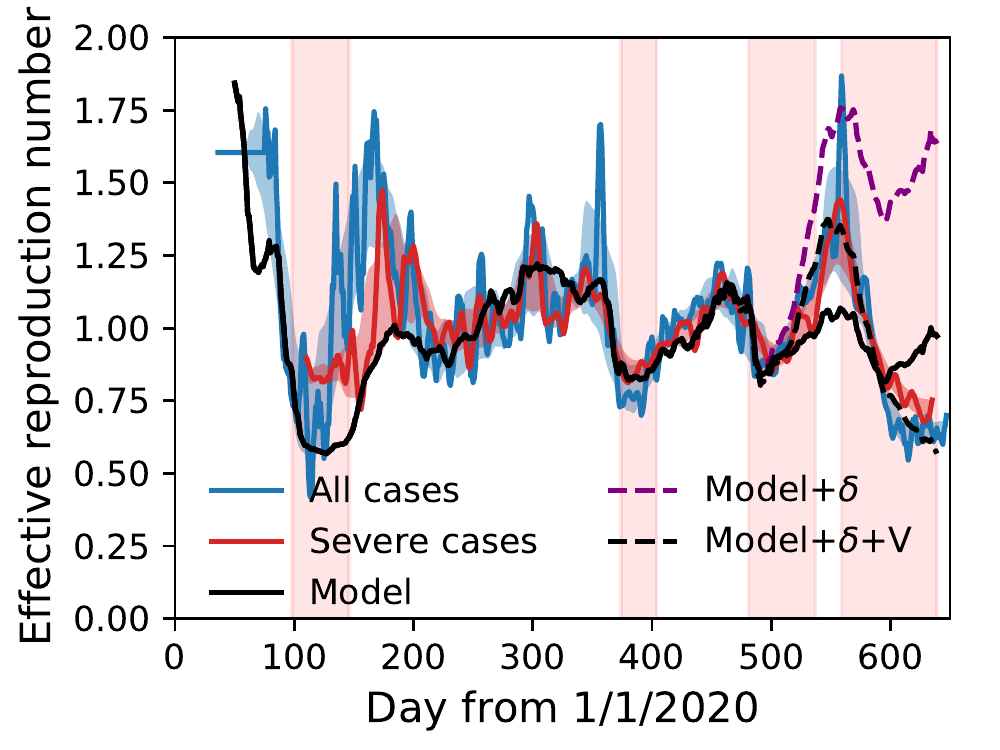}
        \caption{Effective reproduction number of all infections and severe cases in Tokyo metropolitan area compared to the model result. The effective reproduction number is plotted with a delay of 14 and 26 days for all and severe cases, respectively. Parameter fitting is done in the span $200\leq t < 500$. `Model' indicates the prediction from human mobility alone, `Model+$\delta$' includes the effect of the Delta variant, and `Model+$\delta$+V' additionally includes the effect of the vaccination.}
        \label{fig:effective_reproduction_number_fit_DV}
%    \end{minipage}
\end{figure}

We determine the other parameters of the Delta variant and the vaccination after $t=500$ by fitting the epidemiological data as $r_\delta=1.68 \pm 0.03$, $t_\delta=530 \pm 2$, and $C_V=0.99 \pm 0.02$, where the model equation is finally
\begin{equation}
    R_e^{\mathrm{data}}(t+\Delta T)=\left[1+(r_\delta-1)S(\gamma\log(r_\delta)\cdot(t-t_\delta))\right](1 - C_V R_V(t))\frac{1}{S(t)}\sum_{a}\beta_a M_a(t). 
\end{equation}
Figure \ref{fig:effective_reproduction_number_fit_DV} implies that the effects of the Delta variant and the vaccination are fully explained. The values of the parameters, effective reproduction number ratio of the Delta variant, and vaccination's effectiveness are comparable to those reported previously \cite{Covid:DeltaRt01,Covid:DeltaRt02,Covid:VacEff01,Covid:VacEff02,Covid:VacEff03}.

\subsection*{Components of effective reproduction number}

\begin{figure}[htbp]
%    \begin{minipage}[b]{0.5\linewidth}
        \centering
            \includegraphics[width=8cm]{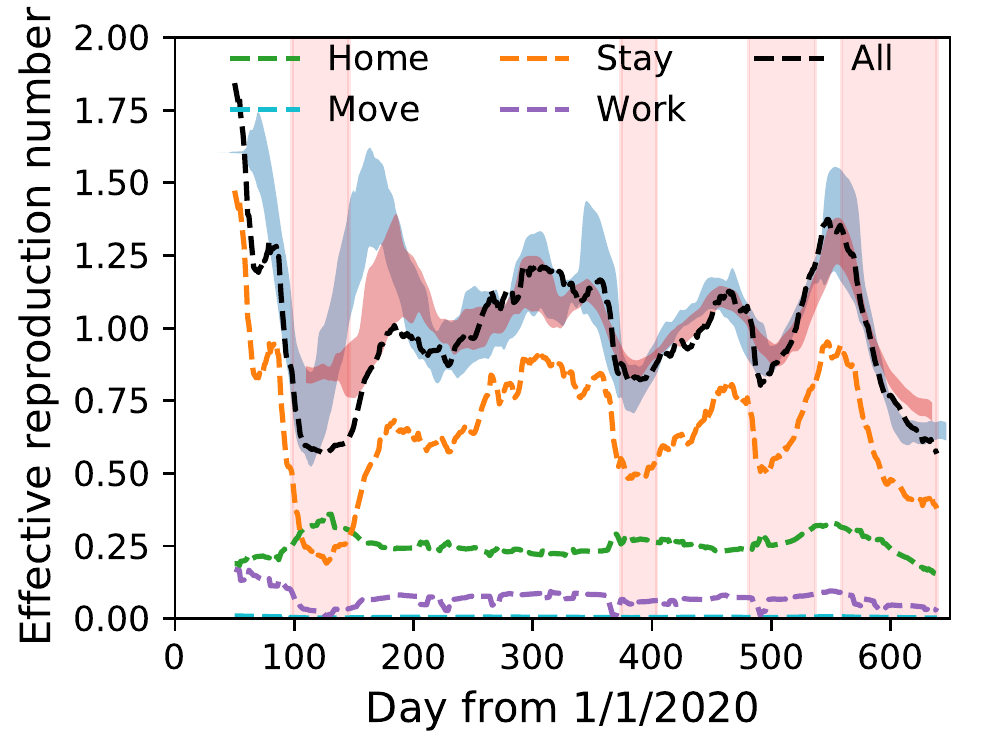}
        \caption{Components of effective reproduction number in the model. Empirical data are indicated by blue and red bands for all infections and severe cases, respectively. The stay-out activity gives the largest component in most time series. \\\quad\\\quad\\\quad\\\quad\\\quad }
        \label{fig:effective_reproduction_number_fit_HMSW_DV}
%    \end{minipage}
\end{figure}

\begin{figure}[htbp]
%    \begin{minipage}[b]{0.5\linewidth}
        \centering
            \includegraphics[width=8cm]{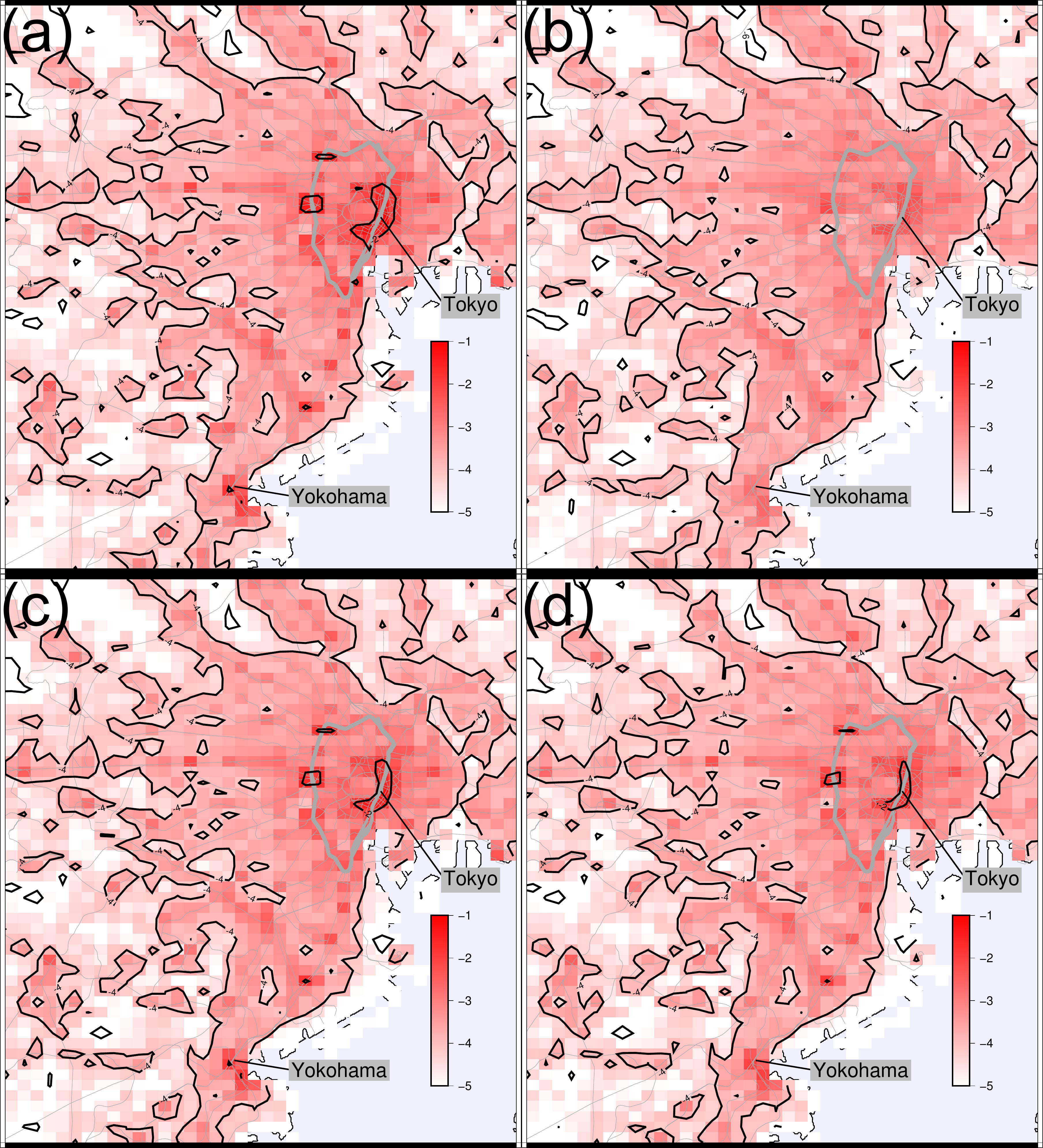}
        \caption{Effective reproduction number map of Tokyo metropolitan area (a) two months before 1st SoE, (b) during 1st SoE, (c) two months before 2nd SoE, and (d) during 2nd SoE. The Yamanote line is indicated by a bold grey closed curve, and other railways are indicated by fine grey curves. The colour represents the values on a log-10 scale. The sum of all regions is the overall effective reproduction number. High-risk zones are concentrated around Tokyo station and the Yamanote line. }
        \label{fig:effective_reproduction_number_map}
%    \end{minipage}
\end{figure}

\begin{figure}[htbp]
%    \begin{minipage}[b]{0.5\linewidth}
        \centering
            \includegraphics[width=8cm]{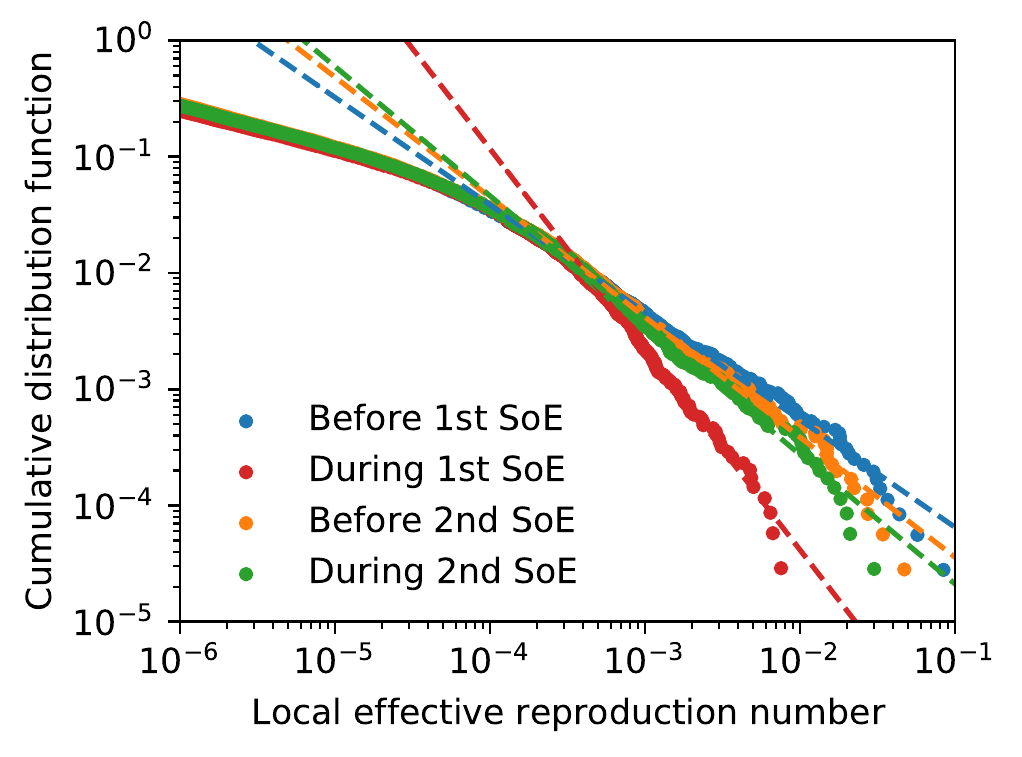}
        \caption{Cumulative distribution function of effective reproduction number at each location on a log-log scale (a) two months before 1st SoE, (b) during 1st SoE, (c) two months before 2nd SoE, and (d) during 2nd SoE. Power-law fitting functions are plotted as dotted lines. The exponents are (a) -0.92, (b) -1.72, (c) -1.03, and (d) -1.11. The sum of the distribution is the total effective reproduction number.}
        \label{fig:effective_reproduction_number_map_CDF}
%    \end{minipage}
\end{figure}

The effective reproduction number for each activity, location, and person was investigated. Fig. \ref{fig:effective_reproduction_number_fit_HMSW_DV} shows the result for each activity: the sum equals the total effective reproduction number. The stay-out activity dominates the change in the whole effective reproduction number because the parameter for stay-out is 28 and 55 times larger than that for home and move/work, respectively. 
The effective reproduction number at each location $m$ is defined by the partial sum of Eq. (\ref{Rp_by_Moment}) 
\begin{equation}
    R_e(m,t)=\frac{1}{S(t)}\sum_{a}\beta_a \gamma^{-1}\sum_{\tau\in t}\left(S_{a m\tau}(t)\right)^2,
\end{equation}
as shown in Fig. \ref{fig:effective_reproduction_number_map} (a) two months just before the 1st SoE, (b) during the 1st SoE, (c) two months just before the 2nd SoE, and (d) during the 2nd SoE. The movie of the effective reproduction number map each day is provided as Supplementary movie 1, where a raw value is taken instead of a 7-day moving average to calculate the population moments. The sum over the Tokyo metropolitan area gives the total effective reproduction number. High-risk zones are concentrated in downtown Tokyo. Fig. \ref{fig:effective_reproduction_number_map_CDF} shows the cumulative distribution. This figure shows that the distribution is almost the same for low-risk regions, and a power law approximates it for high-risk regions whose exponents are (a) -0.93, (b) -1.73, (c) -1.03, and (d) -1.12. Exponents close to -1 imply that the highest-risk regions dominantly affect the total effective reproduction number. In fact, the top-20 highest-risk 1 km-squares in the Tokyo metropolitan area ($36,898\mathrm{km}^2$) contribute (a) 40\%, (b) 13\%, (c) 32\%, and (d) 27\% of the total effective reproduction number. Consequently, restrictions in the highest-risk downtown regions effectively suppress infections. For example, the total effective reproduction number is reduced by 17\%, 25\%, and 36\% if the infection rate is suppressed to 10\% in the top-5, -10, and -20 highest-risk 1 km-squares in period (a), respectively. With regard to real restrictions, the 1st SoE successfully reduced the population density in the downtown region. The SoEs reduced the exponents; however, the change amount of the exponent before and after SoEs also decreased.

\begin{figure}[htbp]
%    \begin{minipage}[b]{0.5\linewidth}
        \centering
            \includegraphics[width=8cm]{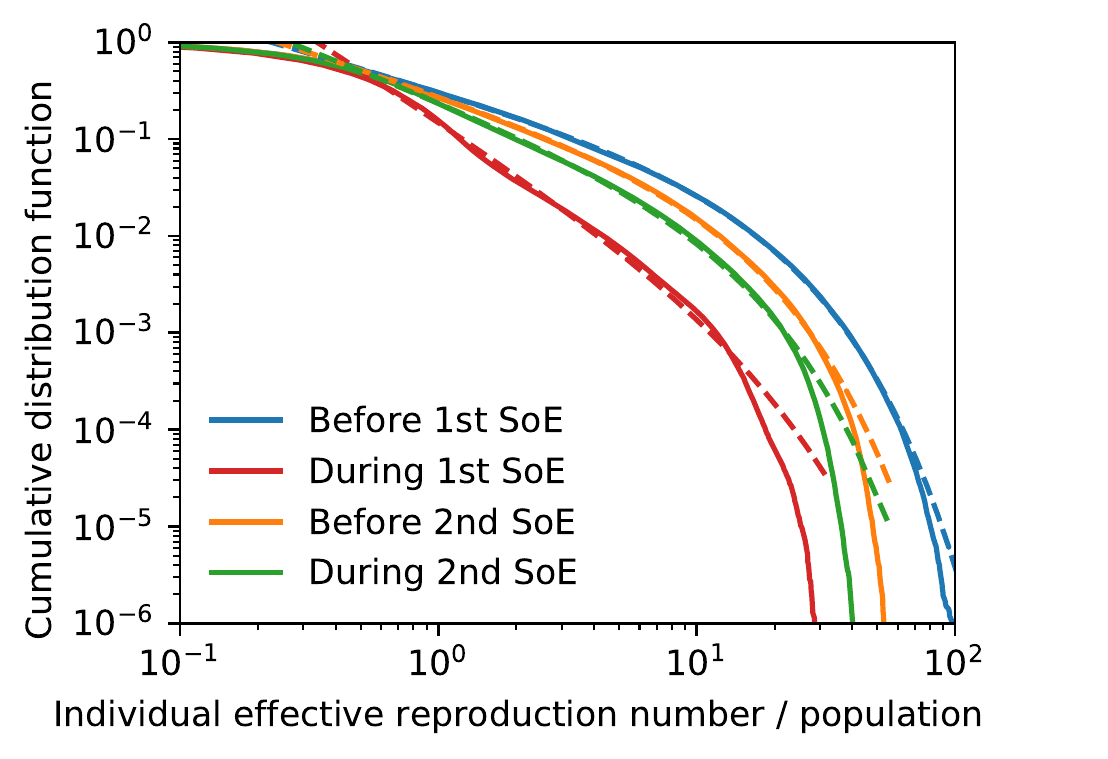}
        \caption{Cumulative distribution function of individual effective reproduction number on a log-log scale (a) two months before 1st SoE, (b) during 1st SoE, (c) two months before 2nd SoE, and (d) during 2nd SoE. Fitting functions are plotted as dotted lines: (a) $0.33 r^{-0.76}\exp(-0.078r)$, (b) $0.16 r^{-1.72}\exp(-0.080r)$, (c) $0.30 r^{-0.86}\exp(-0.105r)$, and (d) $0.26 r^{-1.03}\exp(-0.108r)$, where $R_e^{(i)}$ is denoted by $r$. The mean of the distribution is the effective reproduction number. }
        \label{fig:effective_reproduction_number_each_person}
%    \end{minipage}
\end{figure}

The effective reproduction number for each infected person \cite{IndividualRt01,IndividualRt02} is defined by the mean number of people to whom they would spread the infection; the average for all people gives the overall effective reproduction number. We calculate it using the time series of the activity and the location in a day by the following relation: 
\begin{equation}
    R_e^{(i)}(t)=\sum_{\tau}\beta_{a_i(\tau)} S_{a_i(\tau)m_i(\tau)\tau}(t), 
\end{equation}
where $\tau$ is the time label with 15 min intervals, and $a_i(\tau)$ and $m_i(\tau)$ represent the history of person $i$. We refer to this value as the GPS-based individual effective reproduction number. Fig. \ref{fig:effective_reproduction_number_each_person} shows the cumulative distribution function (CDF) of $R_e^{(i)}(t)$, where we collect all data for (a) two months just before the 1st SoE, (b) during the 1st SoE, (c) two months just before the 2nd SoE, and (d) during the 2nd SoE. The distribution is approximated by truncated power laws. The exponents between -1 and 0 in (a) and (c) clearly indicate that significant cluster infections or superspreaders dominate the total infection. In fact, the top 10\% of people contribute to (a) 58\%, (b) 38\%, (c) 54\%, and (d) 50\% of the overall effective reproduction number. A comparison before and during SoEs reveals that the cut-off does not change while the exponent is decreased. This implies that the effects of the SoE are represented by the reduction of the exponent in the CDF power law; it results in the suppression of the overall effective reproduction number. The cut-off corresponds to the largest infection clusters or superspreaders, which was not suddenly changed by SoEs; however, we observe that it decreased slowly from the 1st to the 2nd SoE.

%$200 \leq t < 500$. The distribution obeys a power law whose exponent is -1.12. The exponent close to 1 indicates that the variance of the infection spread diverges and the cutoff of the power-law dominates the mean. This implies that the significant cluster infection dominates the total infection. 

%TODO: personal Rt Fig



\section*{Discussion}

In this study, we verified human-activity-dependent COVID-19 infection rates using smartphone GPS data. We classified human activity patterns into four types, `home', `move', `stay-out', and `work', and estimated the number of social contacts for each daily activity in the Tokyo metropolitan area. Then, we derived an equation from the classical SIR model for GPS data with activity information to be used directly. The model successfully predicted effective reproduction numbers for future reporting. We demonstrated that infection risk is the highest when the people are not at home or work or not moving. By quantitatively understanding the effect of human mobility on infection spread, we distinguished the impact of the Delta variant and vaccination. Furthermore, we derived formulas that divide the effective reproduction number into the contributions from each location or individual. These formulas enabled us to observe the distributions of infection risk. As applications, we present an effective reproduction number map and GPS-based individual effective reproduction numbers, whose distributions obey the power law or truncated power law. 

The model provides a comprehensive understanding of infection spread of epidemics. A previous research by T. Yabe et al. \cite{Covid:NPITokyo01} investigated the COVID-19 spread in the Tokyo metropolitan area during 2020 in detail and discovered the nonlinear relation between the effective reproduction number and the contact index, the number of social contacts among people not in their homes. The nonlinear relation is explained by the change in the ratio of the population moment of stay-out to that of work and move. Another previous research conducted by S. Rüdiger et al. \cite{Covid:GPS02} shows that the non-uniformity in the infective contact network has an important role as well as the total number of social contacts. The difference from this study in approach is the origin of the non-uniformity in social contacts: the non-uniform distribution of the infectious people in their study (c.f. it can break the assumption (A3)), and the type of the social contacts in this study. 

The assumptions (A1)-(A4) determines the limitation of our model, but relaxing some of them can lead to a new application or complex model for epidemics. 
%If we do not adopt the assumption (A1), the effective reproduction number is decreased to the ratio $(R(t)/N)^2$. This is because the population moment is the square of the susceptible population. 
If we do not adopt the assumption (A1), the effective reproduction number is decreased to the ratio $(1-R(t)/N)$. This is because the possibility of infection in each location decreases to that same ratio. 
For the assumption (A3), the non-uniformity in the infective contact network is considered. Although it is impossible to track each person across the days in the GPS dataset for privacy protection, preferences of people with high infection risk could be assumed or observed in another dataset to estimate the spatial distribution of infected people.

\section*{Methods}

\subsection*{Epidemiological data}

COVID-19 epidemiological data were provided by the Ministry of Health, Labour, and Welfare in Japan \cite{Gov:JapanMHLW}. The data consist of the number of new infection cases and severe cases from Feb. 2020 and May 2020, respectively, to Oct. 2021. We took a 7-day moving average to remove the periodicity of epidemiological data in a week. The moving average was taken from 6 days prior to the target date. Vaccination statistics were obtained from the Prime Minister of Japan and His Cabinet \cite{Gov:JapanKantei01}.

\subsection*{GPS data}

GPS data were provided by Agoop Corp, Japan \cite{Agoop}. They include the location of smartphones with a user ID, time, latitude/longitude, and home/work city from 1 Jan. 2020 to 30 Sep. 2021. The median accuracy of the latitude/longitude data was 10 m. The provider restricts the data resolution for privacy protection if users are close to their home, where the latitude/longitude is fixed to the centre of the home 1 km-square. 

We pre-processed the original data to obtain 15 min interval data using linear interpolation, discarding data that cannot be interpolated sufficiently. The timestamps were every 15 min from 05:00 to 24:00. Data for midnight were discarded because most people stay in bed and do not spread infection. The home/work city data were converted into the home/work 1 km-square data. A home square is a 1 km-square where the user stays at 05:00 in the home city, and a work square is where the user visits continuously for at least 5 h in a day in the work city. Data for iOS smartphones were discarded because they are not gathered if the user is not moving due to the iOS specification, making staying at home undetectable. The converted data cover approximately 0.4 million users in Japan. We investigated the population corresponding to a GPS point to renormalize the GPS data to the actual population distribution. We counted the users at 05:00 for each home prefecture defined by their home city's prefecture for each day. The effective population of one GPS data point in a prefecture was calculated as the ratio of the actual population in the prefecture in Oct. 2019 \cite{Pop2019} to the number of users counted. At the beginning of 2020, the typical values ranged from 250 in metropolitan areas to 500 in the countryside.

\section*{Acknowledgements}
We thank Kenta Yamada, Yukie Sano, Takashi Shimada, and Takahiro Nishi for the helpful discussions. We thank Agoop for providing the GPS datasets. This work was supported by the Tokyo Tech World Research Hub Initiative (WRHI) Program of the Institute of Innovative Research, Tokyo Institute of Technology.

\section*{Data availability}
Our data cannot be open to public, but the same data can be purchased from a Japanese private company, Agoop, which sells "Point-based flow population data". 

\section*{Author contributions}
M.T. was the project leader and directed the entire research plan and writing of the manuscript. Y.S. pre-processed the raw GPS data and appended the activity information. J.O. analysed the pre-processed GPS data, developed the models, performed the numerical calculations, and wrote the manuscript. H.T. improved the data analysis and modelling methods and revised the manuscript.

\section*{Funding}
This work was supported by a Grant-in-Aid for Scientific Research (B) (grant number 18H01656). The authors thank the Tokyo Tech World Research Hub Initiative (WRHI) Program of the Institute of Innovative Research, Tokyo Institute of Technology, for financial support. 

\section*{Competing interests}
The authors declare no competing interests.

\bibliography{main}

\end{document}

\flushbottom
%\maketitle
\thispagestyle{empty}

\renewcommand{\figurename}{Supplementary Figure}

\subsection*{Supplementary note 1 | Estimation of population moment using 1km-square data. }
\begin{figure}[htbp]
    \begin{minipage}[b]{1.00\linewidth}
        \centering
            \includegraphics[width=8cm]{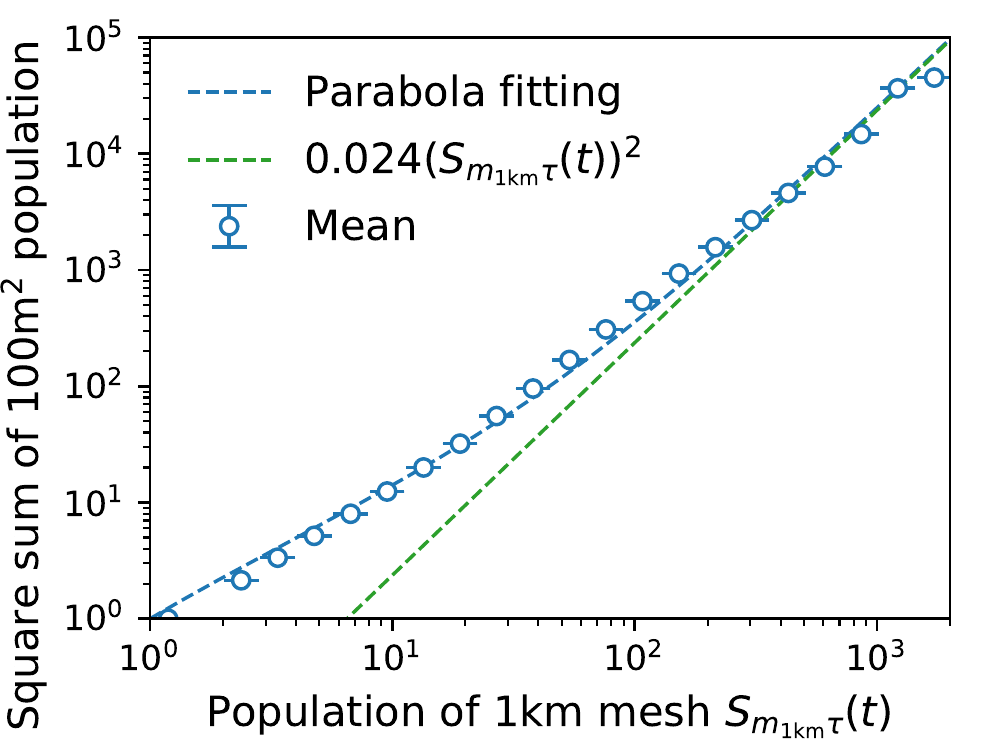}
        \caption{Partial sum of the population moment conditioned by the 1km population without the home activity in Japan. The error bars show the standard deviation of the mean. }
        \label{fig:100mMomentAsAFunctionOf1kmMoment}
    \end{minipage}
\end{figure}

We consider the case of using the 100m-squares instead of the 1km-squares because the social contact is still inhomogeneous within the 1km-squares. We estimate the population moment $M^\mathrm{100m}_a(t)$, defined by the 100m-square data of GPS, from the 1km-square data because the GPS data does not have enough resolution in home squares for privacy protection and enough user number for the data limitation. 
Here we think of 100m-squares labeled by $m_\mathrm{100m}$ within a 1km-square marked by $m_\mathrm{1km}$, and a partial sum $\sum_{m_\mathrm{100m}\in m_\mathrm{1km}}\left(S_{a m_\mathrm{100m}\tau}(t)\right)^2$. We approximate it as a function of the population of 1km-square $S_{am_\mathrm{1km}\tau}(t)$. 
In Fig. \ref{fig:100mMomentAsAFunctionOf1kmMoment}, for all 1km-square in Japan, we plot the partial sum of the population moment $\sum_{m_\mathrm{100m}\in m_\mathrm{1km}}\left(S_{m_\mathrm{100m}\tau}(t)\right)^2$ conditioned by the 1km population $S_{m_\mathrm{1km}\tau}(t)$, where activities do not condition them, and the home activity is removed because of its low resolution. 
Here the 1km population is typically over several hundred if a GPS point is observed. 
This means that the partial sum is approximated by
\begin{equation}
    \sum_{m_\mathrm{100m}\in m_\mathrm{1km}}\left(S_{a m_\mathrm{100m}\tau}(t)\right)^2 \simeq 0.024 \left(S_{am_\mathrm{1km}\tau}(t)\right)^2, 
\end{equation}
where 
\begin{equation}
    \sum_{m_\mathrm{100m}\in m_\mathrm{1km}}S_{a m_\mathrm{100m}\tau}(t) = S_{am_\mathrm{1km}\tau}(t). 
\end{equation}
If the population is homogeneous in the 1km-square, the coefficient must be 0.01, not 0.024. The population inhomogeneity in 1km-squares makes the coefficients 2.4 times more significant. The population moment is described by 
\begin{equation}
    M^\mathrm{100m}_a(t)=\gamma^{-1}\sum_{m_\mathrm{1km}\in A,\tau \in t}\left[\sum_{m_\mathrm{100m}\in m_\mathrm{1km}}\left(S_{a m_\mathrm{100m}\tau}(t)\right)^2\right] \simeq 0.024 \gamma^{-1} \sum_{m_\mathrm{1km}\in A,\tau \in t}  \left(S_{am_\mathrm{1km}\tau}(t)\right)^2=0.024 M_a(t). 
\end{equation}
Therefore the infection rate $\beta^\mathrm{100m}_a$ using the 100m-squares is calculated as 
\begin{equation}
    \beta^\mathrm{100m}_a = \beta_a / 0.024,
\end{equation}
where $\beta_a$ is the infection rate in the 1km-square case.

